\documentclass[%
reprint,
 amsmath,amssymb,
 aps,
floatfix,
]{revtex4-2}

\usepackage{color}
\usepackage{graphicx}
\usepackage{dcolumn}
\usepackage{bm}
\usepackage{relsize}

\begin{document}


\title{Glass transition temperature of thin polymer films}

\author{Hsiao-Ping Hsu}\email{hsu@mpip-mainz.mpg.de}
\author{Kurt Kremer}\email{kremer@mpip-mainz.mpg.de}
\affiliation{Max-Planck-Institut f\"ur Polymerforschung, Ackermannweg 10, 55128, Mainz, Germany}

\begin{abstract}
 The glass transition temperature and its connection to statistical 
 properties of confined and free-standing polymer films of varying thickness containing unentangled to highly entangled bead-spring chains are studied by molecular dynamics simulations. 
 For confined films, perfect scaling of the thickness-dependent end-to-end distance and radius of gyrations normalized to their bulk values in the directions parallel and perpendicular to the surfaces is obtained. Particularly, 
 the reduced end-to-end distance in the perpendicular direction is very well described by the extended Silberberg model. For bulk polymer melts, the relation between chain length and $T_g$ 
 follows the Fox-Flory equation while $T_g$ for a given film thickness is almost independent of chain length. For films, $T_g$ decreases and is well described by Keddie's formula, where the reduction is more pronounced for free-standing films. For the present model, $T_g$ begins to deviate from bulk $T_g$ at the characteristic film thickness, where the average bond orientation becomes anisotropic and the entanglement density decreases. 
\end{abstract}


\date{\today}
\maketitle

In a majority of applications (commodity) amorphous polymeric materials are in the glassy state. Because of that the glass transition temperature $T_g$ region is of crucial importance~\cite{Hrushikesh2017}.  Furthermore, when an amorphous polymer in the liquid state is cooled towards 
$T_g$, the viscosity dramatically increases in a non-Arrhenius way~\cite{Boyd1994,Angell1995,Berthier2011}. 
Despite a huge body of fundamental and applied research the glass transition in general and of polymers in particular~\cite{Li2015,Napolitano2017,McKenna2022} 
 remains to be one of the grand challenges in condensed matter physics. The variable polymer chain length provides an additional parameter for systematic studies compared to low molecular weight glass formers~\cite{Freed2011,Baker2022}.  
The special role polymers can play for a detailed view on the glass transition is illustrated in a recent comprehensive study of the temperature-dependent relaxation dynamics of different polymers on chain length, chain flexibility, and chemical properties 
in Ref.~\cite{Baker2022}. Similarly, Xu et al.~\cite{Xu2020} employ extensive molecular dynamics (MD) simulations based on the model of Ref.~\cite{Hsu2019} to investigate the role of chain stiffness on the glass transition.  Experimentally, typical ways to determine $T_g$ of polymers are by differential scanning calorimetry ({DSC)}~\cite{Mathot1994}, or by 
thermo mechanical analysis (TMA)~\cite{Bird1987}, where different ways to measure $T_g$ can lead to deviating results~\cite{Baker2022}. This holds for both simple, i.e., low molecular weight, glass formers as well as polymers. However, despite decades of research, the nature of the glass transition is still not fully understood~\cite{Angell1988,Angell1991,Binder2005,Baschnagel2005,Stillinger2013,Ediger2014,Schoenholz2017,Mckenna2020,McKenna2022,Berthier2023}. 
These studies mainly refer to bulk systems. Free-standing or supported thin films pose even more challenges and have lead to - partially - contradictory claims in the literature~\cite{Alcoutlabi2005,Serghei2006,Erber2010,FKremer2014,Ma2020}. 
While  mostly it is reported that 
$T_g$ is reduced for very thin films, in some cases depending on chemistry and substrate even an increase was observed~\cite{Mattsson2000,Erber2010,Vogt2018,Keddie1994}. These works stress the relevance of sample preparation and surface coupling or even environmental conditions, leading to questions concerning comparability and equilibration of such sub micron thick films. Due to confinement and surface interaction, significant impact on dynamic and structural polymer properties is observed~\cite{Silberberg1982,Mueller2002,Cavallo2005,Baschnagel2005,Batistakis2012,Sussman2014}. For instance, in-plain chain extensions are only weakly affected by confinement while in the perpendicular direction, the chain extension reduces gradually from the bulk value with decreasing film thickness~\cite{Silberberg1982,Cavallo2005,Sussman2014,Garcia2018}, also leading to chain mobility modifications~\cite{Sussman2014,Garcia2018}, the latter depending on the distance to the surface. Especially for very long, highly entangled polymers these problems become very complex. The aim of this letter is to systematically investigate the dependence of $T_g$ of confined and free-standing films on chain length and film thickness.
 
Taking all these interconnected problems, computer simulations of thin polymer films offer insight under perfectly controlled conditions. Full time dependent coordinates of all particles are available, allowing to investigate the structure and molecular motion (viscosity) and, the glass transition under a variety of situations.~\cite{Xu2020,Hsu2019,Binder2005,Paul2007,Barrat2010}. 
So far however, most computational studies on polymer films focus on short
and unentangled chains~\cite{Barrat2010}, because computing time for equilibration rises dramatically with chain length and systems complexity increases. 

Recently, we have developed an efficient hierarchical methodology to equilibrate highly entangled melts of long polymer chains~\cite{Zhang2014,zhang2019} and extended this to confined and free-standing polymer films~\cite{Hsu2020c}. 
The required computer time 
scales linearly with system size, independent of chain length. While not restricted to standard bead spring chains, we here employ this approach to weakly semiflexible bead spring chains~\cite{Moreira2015,Hsu2016} where the entanglement length $N_e = 28$ beads 
at the standard density of $\rho_0=0.85\sigma^{-3}$. Reference temperature for equilibration and cooling is $T=1 \epsilon/k_B$. Throughout the whole paper Lennard Jones (LJ) units of length ($\sigma$), energy ($\epsilon$,) and time ($\tau$) are used. 


For equilibration and structural properties at $T=1\epsilon/k_B$ we use the mentioned standard semiflexible bead spring model. This model, here referred to as Model I is widely used in the literature and its properties are well documented~\cite{Kremer1990,Kremer1992,Baschnagel2005,Moreira2015,Colmenero2015,Baschnagel2016,Michieletto2017,Svaneborg2020,Raffaelli2020,Abdelbar2023}. Unfortunately this cannot be used to study free surfaces and the glass transition. Model I only contains repulsive non-bonded interactions and displays an atypical chain stretching upon cooling. To correct for that we have developed a variant, Model II, with attractive nonbonded interactions, allowing for free surfaces, and a modified bond angular potential. Details of the parametrization are chosen such that at $T=1 \epsilon/k_B$ and $\rho = 0.85 \sigma^{-3}$ chain conformations and bead packing between the two models is indistinguishable. We thus can take advantage of the huge body of available data of Model I.  A detailed comparison of the models is given in the supporting material (SM)~\cite{SM2023}. This new Model II captures major features of glass-forming polymers, e.g.~that the viscosity and relaxation time dramatically increase in a non-Arrhenius way close to $T_g$~\cite{Hsu2019,Xu2020a,Xu2020b} as it is characteristic for fragile polymeric glass formers.
For confined polymer films, two confining planar, structureless repulsive walls with a  10-4
LJ potential~\cite{Aoyagi2001,Grest1996,Hsu2020c} are introduced. This is sufficient to investigate generic conformational properties of the films. 

\begin{figure}[t!]
\begin{center}
\includegraphics[width=0.30\textwidth,angle=0]{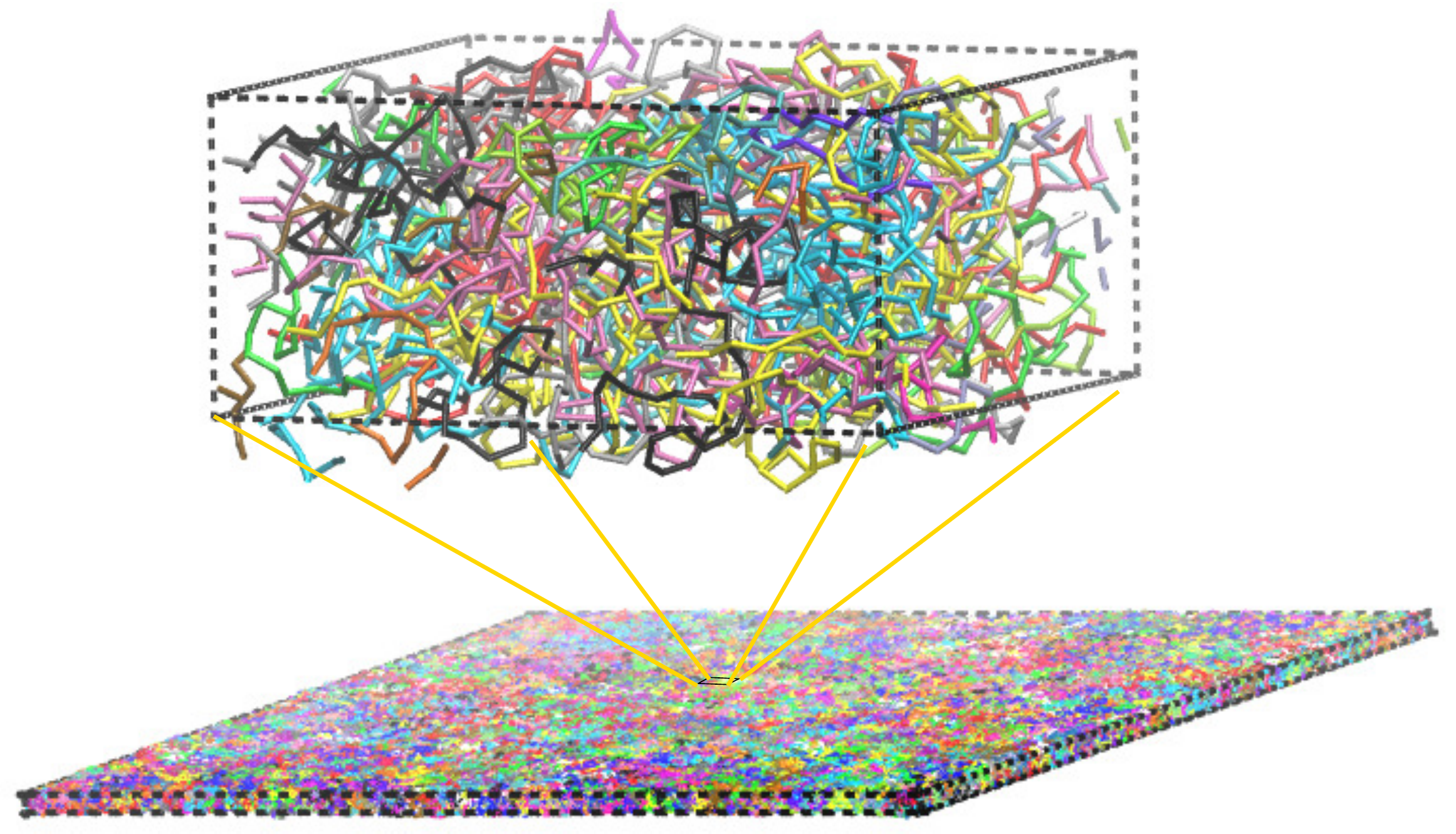}
\caption{Snapshot of a configuration of fully equilibrated free-standing film 
($n_c=1000$, $N=2000\approx 72 N_e$) of $h \approx 9.0\sigma \approx 0.3 R_g^{(0)}(N)$ at $T=1.0\epsilon/k_B$. The dashed box is shown for better  visualization.}
\label{fig-free-standingfilm}
\end{center}
\end{figure}

The ESPResSo++ package~\cite{Espressopp,Espressopp20} is used to perform 
MD simulations with Langevin thermostat in the NVT and NPT ensemble, with a Hoover barostat for the latter.  Our polymer melts contain $n_c$ chains of $N$ monomers ($n_cN=10^5$ for $N\le 100$, and $n_c=1000$ for $500 \le N \le 2000$), ranging from unentangled ($N<N_e$) to highly entangled ($N \gg N_e$) films. 
Film thicknesses $h$ range from thick ($h>R_g^{(0)}$) to very thin ($h<R_g^{(0)}$) films, $R_g^{(0)}$ being the root mean square (rms) radius of gyration for bulk melt chains. The smallest $h \approx 9.0 \sigma<2d_T$ (see Fig.~\ref{fig-free-standingfilm}), $d_T=5.02\sigma$ being the tube diameter~\cite{Hsu2016}. 
In all cases $h$ is measured along the $z$ direction  according to the concept of the Gibbs dividing surface that has been applied to identify the interface between two different phases~\cite{Hansen2013,Hsu2020c}, while periodic boundary conditions are applied in the $x$ and $y$ directions, respectively.
All confined polymer melts of short chains $N\le 100$ easily could be generated via a brute-force equilibration similar as in the bulk~\cite{Moreira2015}.

\begin{figure}[t!]
\begin{center}
\includegraphics[width=0.40\textwidth,angle=0]{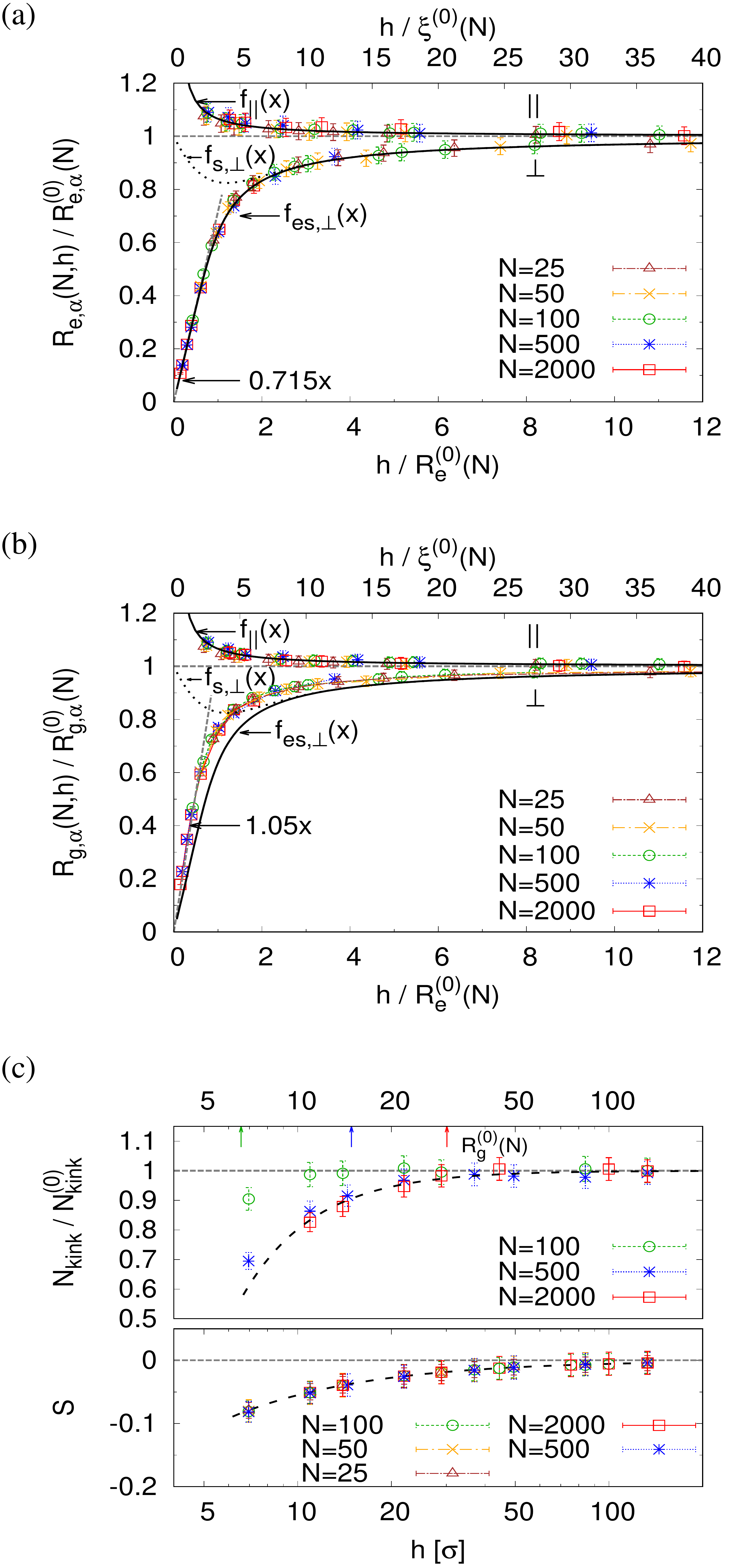}
\caption{(a)(b) Two components of rms end-to-end distance, and radius of gyration rescaled to the bulk value,  $R_{e,\alpha} (N,h)/ R_{e,\alpha}^{(0)}(N)$ (a)
and $R_{g,\alpha} (N,h)/ R_{g,\alpha}^{(0)}(N)$ (b)
in the directions parallel ($\alpha=||$) and perpendicular ($\alpha=\perp$) to the walls, plotted versus $h/\xi^{(0)}(N)$, and $h/R_e^{(0)}(N)$, respectively.
(c) Orientational order parameter $S$, and reduced number of kinks, $N_{\rm kink}/N_{\rm kink}^{(0)}$, plotted 
versus $h$. 
In (a)(b), theoretical predictions $f_{||}(x=h/\xi^{(0)}(N))=1+c_hx^2$ with the fitting parameter $c_h\approx 0.22$, $f_{s,\perp}(x=h/R_e^{(0)})$~\cite{Silberberg1982},  and $f_{es,\perp}(x=h/R_e^{(0)})$~\cite{Sussman2014} (cf.~text) are shown by curves for comparison. In (c), $N_{\rm kink}^{(0)}=4.28(32)$, $22.28(1.31)$, and $91.78(2.92)$ for $N=100$, $500$, and $2000$, respectively, and the dashed lines are drawn to guide the eye. All data are for $T=1.0\epsilon/k_B$. } 
\label{fig-ReRg-confined}
\end{center}
\end{figure}

We first analyze the chain conformations for confined melts at 
$T=1.0 \epsilon/k_B$ in comparison to bulk properties as shown in Fig.~\ref{fig-ReRg-confined}. 
For each component of the 
rms end-to-end distance and radius of gyration we find excellent data collapse onto a universal master curve. This is an additional indication that all confined polymer films are indeed fully equilibrated.  While $R_{e,||}(N,h)$, $R_{g,||}(N,h)$ remain unchanged almost down to 
$h\approx 2R_e^{(0)}$ this is different for the perpendicular component. There we observe a gradual decrease already for much thicker films originating from chains close to the surface. Around $ h \approx R_e^{(0)}$ this turns into a linear decrease with $h$ for $R_{e,\perp}(N,h)$ and ($R_{g,\perp}(N,h)$),  while $R_{e,||}(N,h)$ ($R_{g,||}(N,h)$) weakly increase~\cite{Cavallo2005}. 
Already 1982 Silberberg~\cite{Silberberg1982} argued that chains next to a wall can be treated as unperturbed random walks located with their center at the wall and folded back on one side of the wall. This changes $R_{e,\perp}(N,h)$, however, does not affect the parallel component. This idealized picture works well in the thick film regime down to $h \approx 2R_e^{(0)}$. Sussman et al.~\cite{Sussman2014} extended Silberberg's hypothesis from one wall to two walls, keeping the reflecting boundary conditions in the perpendicular direction and the Gaussian weight factor for each contribution. Their theoretical prediction of $f_{es,\perp}(x=h/R_e^{(0)}(N))$, shown in (SM)~\cite{SM2023}
up to the second-order correction term, is supported by our data covering the thick and thin film regime. For ideal chains $R_e^{(0)}/R_g^{(0)}\approx \sqrt{6}$ while the distribution of $R_g^{(0)}$ 
is not exactly Gaussian anymore. Therefore, our data of $R_{g,\perp}(N,h)$ slightly deviate from $f_{es,\perp}(x=h/R_e^{(0)}(N))$. 
Furthermore, a detailed analysis of our data in the parallel direction reveales that scaling behavior of $R_{e,||}(N,h)$ and $R_{g,||}(N,h)$ is better described by $f_{||}(x)=1+c_h x^2$ instead of $f_{||}(x)=1+c_h x$, with  $x=(h/\xi^{(0)}(N))^{1/2}$ as in Ref.~\cite{Mueller2002}. Here $\xi^{(0)}(N)=R_g^{(0)}(N)/\sqrt{\bar{N}}\approx const$ (see Fig.~S1) is the excluded volume screening length. $\sqrt{\bar{N}}=\rho_0 (R_g^{(0)}(N))^3/N \approx 0.2563 N^{1/2}$ is the degree of interdigitation of different polymers, also called generalized polymerization index~\cite{Mueller2002} in bulk melts.

\begin{figure}[t!]
\begin{center}
\includegraphics[width=0.25\textwidth,angle=270]{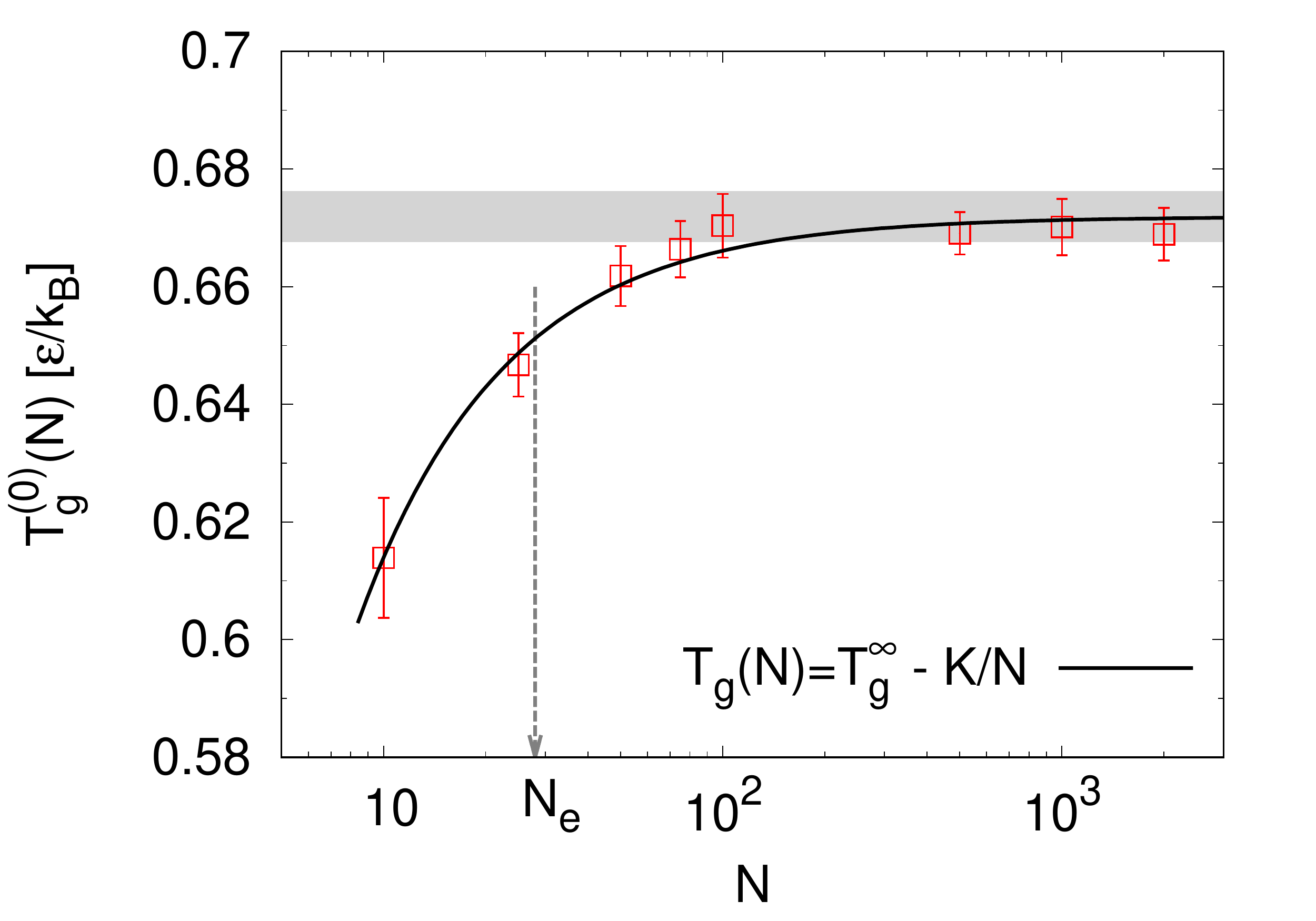}
\caption{Estimates of $T_g^{(0)}(N)$ from the density $\rho(T)$ change, plotted as a function of $N$ at $P=0\epsilon/\sigma^3$ 
for polymer melts in bulk. The Fox-Flory relation~\cite{Fox1950} with $K=0.579(59)$ and $T_g^\infty =0.6718(44)$ is shown by a black curve for comparison. The uncertainty of $T_g^\infty$ is indicated by a shaded gray region.}
\label{fig-Tg-bulk}
\end{center}
\end{figure}

\begin{figure}[t!]
\begin{center}
\includegraphics[width=0.3193\textwidth,angle=0]{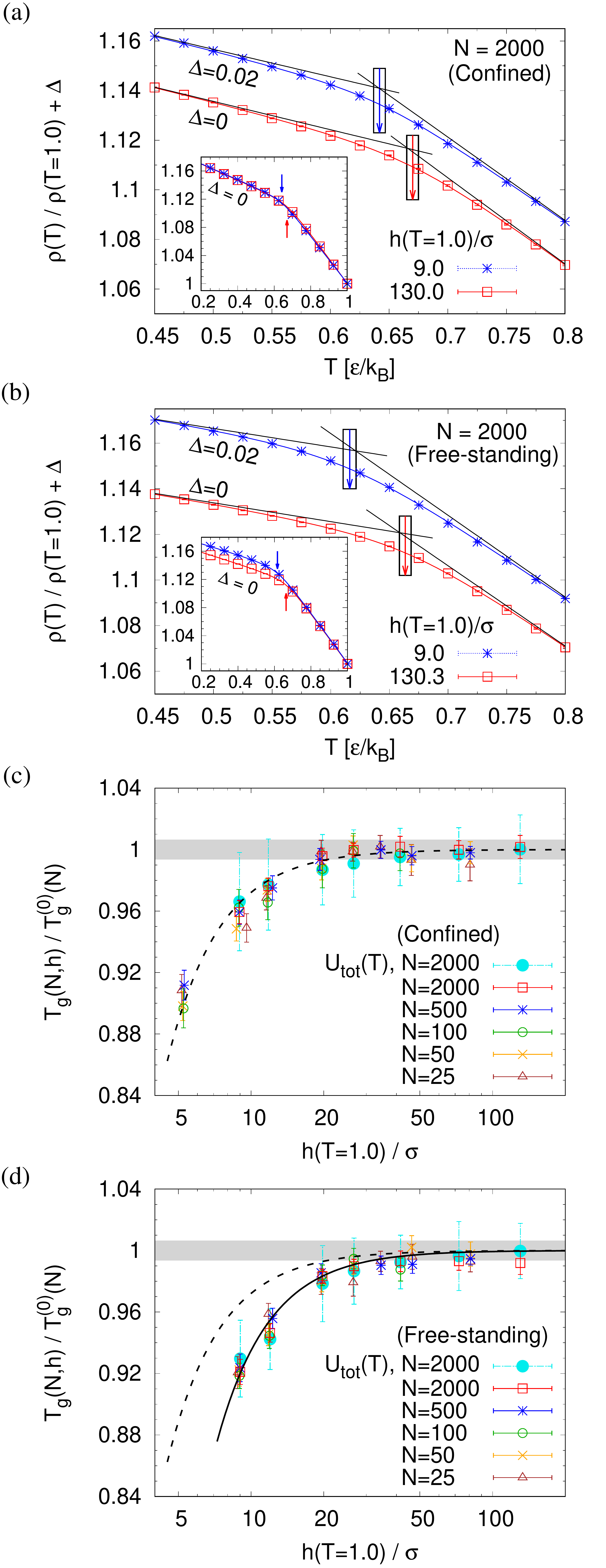}
\caption{(a)(b) Rescaled monomer density $\rho(T)/\rho(T=1)$ in confined (a) and free-standing polymer films (b) of 
$h/R^{(0)}(N)\approx 4.3$ and $0.3$, plotted 
versus $T$ for $N=2000$. The hyperbolic and tangent fits are represented by curves and lines, respectively. Estimates of $T_g(N,h)$ via the hyperbolic fit are indicated by arrows including the uncertainty.
In the inset of (a)(b), full data for $0.2 \le k_BT/\epsilon \le 1.0$ are shown. 
(c)(d) $T_g(N,h)$ rescaled by $T_g^{(0)}(N)$ (see Fig.~\ref{fig-Tg-bulk}) including the error bars, plotted as a function of $h$ for confined (c) and free-standing (d) films. 
The uncertainty of $T_g^{(0)}(N)$ is indicated by a shaded gray region. The formula $T_g(h)/T_g^{(0)}=1-(h_0/h)^\delta$ proposed by Keddie et al.~\cite{Keddie1994} with $\delta=2.0(1)$, $h_0=1.67(23)\sigma$ in (c), and $h_0=2.55(35)\sigma$ in (d) are shown by dashed, and solid curves, respectively. Estimates of $T_g(N,h)$ from 
the potential energy $U_{\rm tot}(T)$ change for $N=2000$ are also shown in (c)(d) for comparison. At $T=1.0\epsilon/k_B$, 
$\sigma^3 \rho(T) \approx 0.854$, $0.858$ for $h/\sigma\approx 130.0$, $9.0$, respectively in (a), and $\sigma^3 \rho(T) \approx 0.853$, and $0.852$ for $h/\sigma \approx 130.3$, $9.0$, respectively in (b).}
\label{fig-Tg-film}
\end{center}
\end{figure}

This change of conformation of confined films also affects the orientational bond distribution in the chains as analyzed by the of the bond orientational order parameter $S=(3\langle \cos \phi \rangle -1)/2$. $\phi$ being the angle between each bond and the ${z}$-axis. Furthermore entanglements will be affected by the change in chain self density. Applying a primitive path analysis~\cite{Everaers2004} 
to obtain entanglement points (significant kinks)~\cite{Hsu2018a,Hsu2018b} along the primitive paths (PPs) reveals this. 
Results of $S$ and the average number of significant kinks $N_{\rm kink}$ as a function of $h$ are shown in Fig.~\ref{fig-ReRg-confined}c. $S$ is independent of $N$ for a given $h$, indicating that the local packing is not affected by $N$. In contrast estimates of $N_{\rm kink}$ for $N=500$ and $2000$ follow a master curve and deviate strongly from that of shorter chains of only a few entanglement lengths ($N=100 \approx 3.6 N_e$).
However, both data sets deviate from the bulk below and around $h=h_c\approx 20.0\sigma$. Obviously, $h_c$  is not related to $R_g^{(0)}(N)$ ($R_e^{(0)}(N)$), differently from the prediction in Ref.~\cite{Sussman2014}.  Chains are less strongly oriented on the scale of bond vectors between monomers 
compared to the scale of the end-to-end vector~\cite{Garcia2018}.

To model free-standing films in vacuum, and to study the glass transition of polymer films, we apply Model II to all confined polymer films shown in Fig.~\ref{fig-ReRg-confined} at $T=1.0\epsilon/k_B$. After a short initialization in the NVT ensemble, 
the confined polymer films are further relaxed for about $13\tau_e$ in a NPT ensemble with fixed wall distance at pressure $P=0.0\epsilon/\sigma^3$. This led to a marginal adjustment of the  lateral extensions of the films. Then, free-standing films are obtained just by removing the wall potential while keeping their lateral dimensions fixed (for more details see Ref.~\cite{Hsu2020c}). 
After that, we perform MD simulations of confined films, and bulk melts in the NPT ensemble and in the NVT ensemble for free-standing films, keeping the dimensions of films constant. 
The component of pressure tensor along the direction perpendicular to the interfaces $P_{zz}$ for all films fluctuates around zero. 

To study the glass transition, we follow exactly the very same cooling protocol as for our 
previous bulk studies of the same polymer model~\cite{Hsu2019,Singh2020,NIC2022}. We apply stepwise cooling~\cite{Buchholz2002} which results in a cooling rate $\Gamma=\Delta T/\Delta t=8.3\times 10^{-7}\epsilon/(k_B\tau)$. The temperature is reduced in steps of $\Delta T=0.025\epsilon/k_B$ from $T=1.0\epsilon/k_B$ to $0.2\epsilon/k_B$ with a relaxation time between each step of $\Delta t=30000\tau \approx 13\tau_e$ ($\tau_e=\tau_0N_e^2 \approx 2266\tau_e$ being the entanglement time~\cite{Hsu2016}), i.e. subchains of the order of $N_e$ can relax easily at higher temperatures close to $T=1.0\epsilon/k_B$. 

To determine $T_g$, we perform a hyperbolic fit~\cite{Patrone2016} to the density $\rho(T)$ change with temperature
$\rho(T)=c-a(T-T_0)-b/2(T-T_0+\sqrt{(T-T_0)^2+4e^f})$ where $c$, $T_0$, $a$, $b$, and $f$ are fitting parameters. Adopting this fit, $T_g$ is either defined by $T_g=T_0$ or the intersection point of two tangents drawn at the high and the low temperature. Both give the same estimate within fluctuations for all systems studied here, see e.g.~Fig.~\ref{fig-Tg-film}ab for confined and free-standing films of two selected film thickness $h$ 
for the longest polymer chains of $N=2000$. 
For comparison, the estimates of $T_g^{(0)}(N)$ of corresponding bulk melts~\cite{Hsu2019,Singh2020,NIC2022} 
are shown in Fig.~\ref{fig-Tg-bulk}.
Similar as in experiment and other simulations, our data are well described by the Fox-Flory relation~\cite{Fox1950,Baker2022,Barrat2010}. $T_g^{(0)}(N)$ decreases with $N$ for $N < 2N_e$ while $T_g^{(0)}(N) \approx T_g^\infty=0.6718(44)$ for $N \gtrsim 2N_e$. 

Fig.~\ref{fig-Tg-film}cd shows the results of $T_g(N,h)$ of confined and free-standing films depending on $h$ and $N$ following the same data analysis as shown in Fig.~\ref{fig-Tg-film}ab.
Obviously, $T_g(N,h)$ covering the range from unentangled to highly entangled chains only very weakly depend on $N$, while the reduction with decreasing $h$ in both confined and free-standing films is clearly observed. Data are well described by $T_g(h)=T_g^{(0)}(1-(h_0/h)^\delta)$~\cite{Keddie1994} with an exponent $\delta=2.0(1)$, consistent with $\delta=1.8(2)$ for the supported PS films~\cite{Keddie1994}. The characteristic length $h_0$ for the free-standing films is about $1.5$ times that for confined films due to the different chain mobility near the surface~\cite{Ediger2014}. 
The deviation of $T_g(N,h)$ from $T_g^{(0)}(N)$ around $h \approx 20\sigma$ fits well to the observed changes of the bond orientation and the reduction of entanglements in Fig.~\ref{fig-ReRg-confined}c. Most notably, the relative depletion of $T_g(N,h)$ seems to be almost independent of chains length, unlike the value of $T_g(N)$ in bulk. Alternatively, one can also determine $T_g$ from the bi-linear fits of the total potential energy $U_{\rm tot}(T)$ change~\cite{Buchholz2002,Zhang2017} (see SM~\cite{SM2023}). Estimates of $T_g$ for $N=2000$ are included in Fig.~\ref{fig-Tg-film}. 
The $T_g$ reduction remains the same, however the uncertainty is much larger.

In summary, we confirm that the dependence of $T_g$ of bulk polymer melts on the chain length $N$ in the range from unentangled to highly entangled can be well described by the Fox-Flory equation. The scaling prediction of the chain extensions in the directions parallel and perpendicular to the surface of confined films are verified in both thick and thin film regimes. We show that $T_g$, $S$ and $N_{\rm kink}$ (for films of highly entangled polymer melts) start to deviate from the bulk value as the effective film thickness $h \lesssim h_c \approx 20\sigma$ related to the intrinsic properties of polymer films while the $T_g$ reduction is stronger for the free-standing films. However, a detailed study of $T_g$ in the layers of polymer films, the distribution of entanglements inside the films, and the mobility of polymer chains near the surface is needed for the further understanding of the $T_g$ reduction.

\textbf{Acknowledgement} We are grateful to Denis Andrienko for a critical reading of the manuscript. This work was supported by the European Research Council under the European Union’s Seventh Framework Programme (FP7/2007-2013)/ERC Grant Agreement No. 340906-MOLPROCOMP. We also gratefully acknowledge the computing time granted by the John von Neumann Institute for Computing (NIC) and provided on the supercomputer JUWELS at the Jülich Supercomputing Centre (JSC), and the Max Planck Computing and Data Facility (MPCDF).


%

\pagebreak
\onecolumngrid

\renewcommand{\thefigure}{S\arabic{figure}}
\setcounter{equation}{0}
\setcounter{figure}{0}
\setcounter{table}{0}
\setcounter{page}{1}
\makeatletter
\renewcommand{\theequation}{S\arabic{equation}}
\renewcommand{\thefigure}{S\arabic{figure}}
\renewcommand{\bibnumfmt}[1]{[S#1]}
\renewcommand{\citenumfont}[1]{S#1}

\begin{center}
\textbf{\large Supplementary Material\\ Glass transition temperature of thin polymer films}
\vskip 0.5truecm
Hsiao-Ping Hsu and Kurt Kremer
\vskip 0.1truecm
{\it Max-Planck-Institut f\"ur Polymerforschung, Ackermannweg 10, 55128, Mainz, Germany} 
\end{center}
\vskip 0.5truecm

\section{Simulation models}
For our simulations, two variants of the standard bead-spring models~\cite{Kremer1990,Kremer1992} are used. In the first model (Model I), which is frequently used and analyzed in detail~\cite{Kremer1990,Kremer1992,Baschnagel2005,Moreira2015,Colmenero2015,Baschnagel2016,Michieletto2017,Svaneborg2020,Raffaelli2020,Abdelbar2023}, a bond-bending potential $U_{\rm BEND}^{\rm (old)}(\theta)=k_\theta(1- \cos \theta)$~\cite{Auhl2003} with $k_\theta=1.5\epsilon$ is added to control the chain stiffness.  Beyond that the FENE and the truncated and shifted LJ potentials, $U_{\rm FENE}(r)$ and $U_{\rm WCA}(r)$, for the linkage and excluded volume interactions between monomers are kept. T=$1.0 \epsilon/k_B$ is the equilibration temperature and reference for all comparisons to bulk behavior. The reference density is $\rho = 0.85 \sigma^{-3}$ as in all above mentioned studies. Two planar, structureless repulsive walls described by the 10-4 LJ potential~\cite{Aoyagi2001,Grest1996,Hsu2020c} are then introduced for studying confined polymer films. This is sufficient to investigate generic conformational properties of thin, highly entangled polymer film confined between two plates. 

\begin{figure}[h!]
\begin{center}
\includegraphics[width=0.28\textwidth,angle=270]{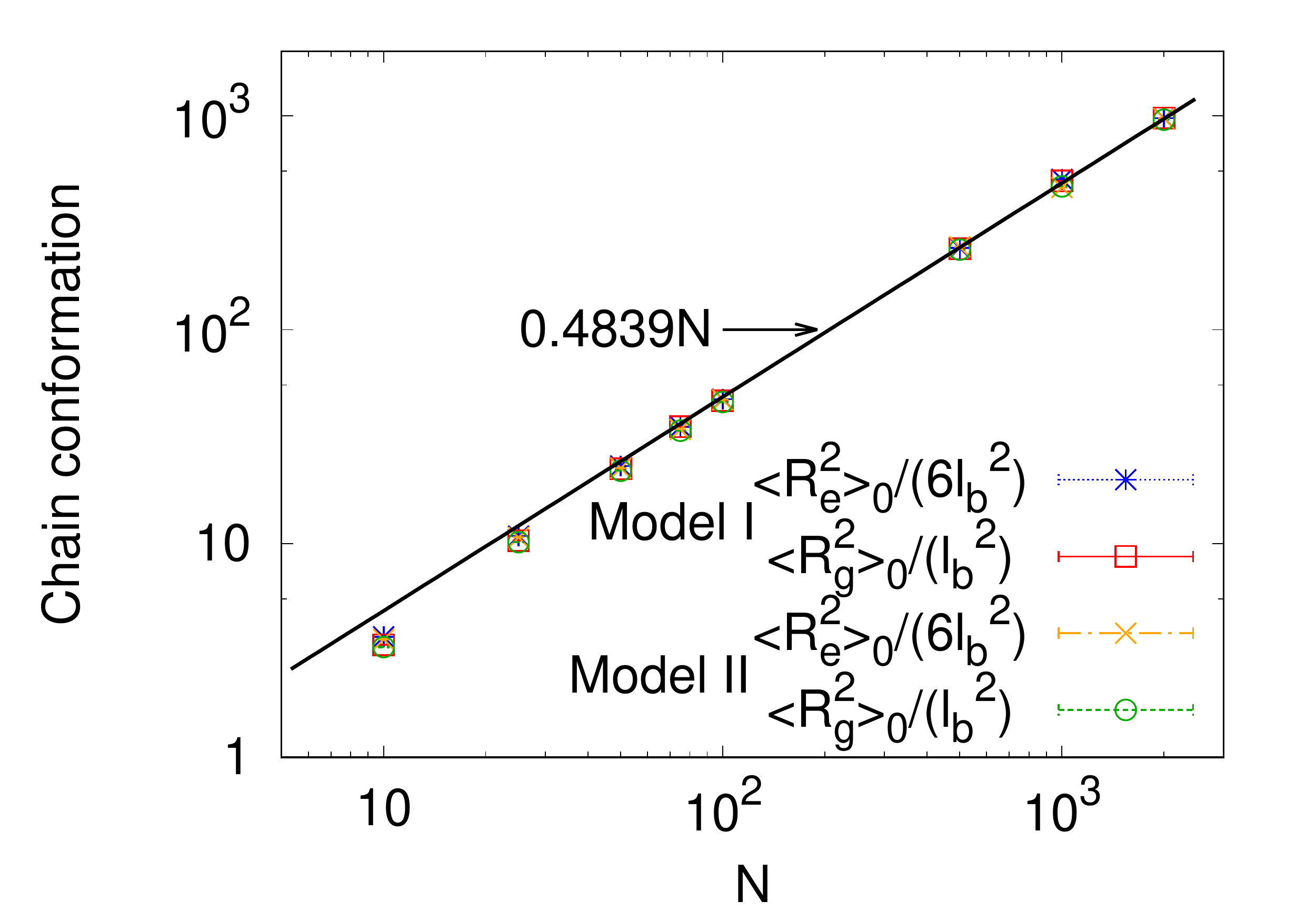}
\caption{Rescaled mean square end-to-end distance $\langle R_e^2 \rangle_0/(6\ell_b^2)$, and radius of gyration $\langle R_g^2 \rangle_0/\ell_b^2$ with the bond length $\ell_b\approx 0.964\sigma$~\cite{Hsu2016} plotted versus chain length $N$ on a log-log scale at $T=1.0\epsilon/k_B$ for polymer melts in bulk, 
described by two variants of bead-spring models, as indicated. The straight line $0.4839(47)N$ gives the best fit to our simulation data. For convenience, we define $R_e^{(0)}(N)=\langle R_e^2(N) \rangle_0^{1/2}$ and $R_g^{(0)}(N)=\langle R_g^2(N) \rangle_0^{1/2}$ in the main text.}
\label{fig-bulk}
\end{center}
\end{figure}

Unfortunately this widely used model is not appropriate to investigate the glass transition and as well as free-standing films. The repulsive bead-bead interaction alone does not allow for free surfaces.  Moreover, the bending potential introduced leads to an artificial chain stretching upon cooling. For that the recently developed new model (Model II) was parameterized, which at T=$1.0 \epsilon/k_B$ and pressure $P=0 \epsilon / \sigma^3$ equilibrates at the bulk melt density $\rho = 0.85 \sigma^{-3}$ and produces - within the error bars -  identical chain conformations and bead packing. Thus it has been shown that it is possible to seamlessly switch from melts of Model I to melts of Model II at T=$1.0 \epsilon/k_B$~\cite{Hsu2019}. In short, the standard bond-bending potential $U_{\rm BEND}^{\rm (old)}(\theta)$ is removed and two additional potential energy terms are introduced. A short-range non-bonded attractive potential $U_{\rm ATT}(r)=\alpha[\cos(\pi(r/r_{\rm cut})^2)-1]$ for $r_{\rm cut}=2^{1/6}\sigma \le r < r_c^a=\sqrt{2}r_{\rm cut}$ is added, where $r_{\rm cut}$ is the cut-off distance for $U_{\rm WCA}(r)$.  $\alpha=0.5145\epsilon$ is determined such that the pressure drops to $P \approx 0\epsilon/\sigma^3$ for bulk long chain polymer melts of $n_c=1000$ chains of $N=2000$ monomers. The standard bond-bending potential $U_{\rm BEND}^{\rm (old)}(\theta)$ is replaced by a new $U_{\rm BEND}(\theta)=-a_\theta \sin^2(b_\theta \theta)$ for $0<\theta<\theta_c=\pi/b_\theta$ such that chain conformations only weakly depend on $T$ as observed in experiments~\cite{Fetters2007}. $a_\theta=4.5\epsilon$ and $b_\theta=1.5$ are fitted to perfectly match results obtained using Model I with $k_\theta=1.5\epsilon$~\cite{Hsu2019} at $T=1.0\epsilon/k_B$. As shown in Fig.~\ref{fig-bulk}, the resulting chain conformations remain the same for bulk melts after switching to Model II, and chains behave as ideal chains. 

\bigskip

\section{Theoretical predictions of chain conformations in confined polymer films}

 The chain conformational change described by the mean square end-to-end distance in the direction perpendicular to the walls ($z$-direction) at a distance $h$ compared to the bulk value is given by
\begin{equation}
      r_{e,\perp}^2(h)=\frac{1}{h}\int_0^h \frac{R_{e,\perp}^2(z^*)}{(R^{(0)}_{e})^2/3} dz^*
\end{equation}
where $z^*$ denotes the $z$-coordinate of one end of chains, and $R_e^{(0)}=\langle R^2_e \rangle_0$ is the root mean square end-to-end distance of chains in a bulk melt. Rescaled $h$ by $R_e^{(0)}$, $f_{s,\perp}(x)$ and $f_{es,\perp}(x)$ in the main text denote 
the master scaling functions of $r_{e,\perp}(x=h/R_e^{(0)})$ predicted by Silberberg~\cite{Silberberg1982} and Sussman et al.~\cite{Sussman2014}, respectively.
The corresponding formulas of $R^2_{e,\perp}(z^*)$ in Ref.~\cite{Silberberg1982,Sussman2014} are given below. 

\subsection{Silberberg prediction in the thick films} 

\subsubsection{Polymer chains in a bulk}

Total number of configurations for chains of $N+1$ monomers starting at $z^*$ is equivalent to chain ending at $z^*$
\begin{equation}
 Q^{\rm tot}_{z^*}=\sum_{z_{N}>z^*} Q(z_{0}=z^*,z_{N})=\sum_{z_{0}<z^*} Q(z_{0},z_{N}=z^*) \,.
\end{equation}
Assuming that chains behave as random walks, the distribution of end-to-end vector of chains follows a Gaussian distribution. One obtains, to a good approximation, 
\begin{equation}
  Q(z_{0},z_{N}) \cong Q^{\rm tot}C e^{-\beta(z_{N}-z_{0})^2} \qquad \textrm{with} \qquad
   \int_{z_0}^{\infty} Q(z_{0},z_{N}) d z_{N}=1
\end{equation}
where $C=2\sqrt{\beta/\pi}\,$. The mean square end-to-end distance in the $z$-direction is given by,
\begin{eqnarray}
  (R^{(0)}_{e,\perp})^2 =\frac{1}{3}(R_e^{(0)})^2&=&\langle (z_N-z_0)^2 \rangle \nonumber \\
 &=& \int_{z_0}^{\infty} (z_N-z_0)^2 Q(z_0,z_N) dz_N  \nonumber \\
  &=& 2 \sqrt{\frac{\beta}{\pi}} \int_{z_0}^\infty (z_N-z_0)^2 e^{-\beta(z_N-z_0)^2}  dz_N \nonumber \\
  &=& \left( 2 \sqrt{\frac{\beta}{\pi}}\right) \left( \frac{1}{4} \sqrt{\frac{\pi}{\beta^3}}\right) = \frac{1}{2\beta} \,.
\end{eqnarray} 

\subsection{Polymer chains near the surface}

The number of configurations of chains starting at $z^*$ in the presence of a wall at $z=0$ is given by
\begin{eqnarray}
  Q^w (z_0=z^*, z_N)&=& Q^{\rm tot} C \left[\int_{z^*}^\infty e^{-\beta(z_N-z^*)^2} dz_N 
    +\int_{-\infty}^{-z^*} e^{-\beta(z_N-z^*)^2} dz_N \right] \nonumber \\
&=& Q^{\rm tot}\left[2-\textrm{erf}(2z^*\sqrt{\beta})\right] \,.
\end{eqnarray}
Similarly, the number of configurations of chains terminating at $z^*$ is given by
\begin{eqnarray}
  Q^w (z_0,z_N=z^*) 
= Q^{\rm tot} C \left[\int_{-z^*}^{z^*}  e^{-\beta(z^*-z_0)^2} dz_0 \right]
= Q^{\rm tot} \textrm{erf}(2z^*\sqrt{\beta}) \,.
\end{eqnarray}
We therefore can write down the formula of $R^2_{e,\perp}(z^*)$ in the $z$-direction for chains of one end located at $z^*$ as follows,
\begin{eqnarray}
 \left(Q^w (z_0=z^*, z_N)+Q^w (z_0,z_N=z^*) \right) R_{e,\perp}^2 (z^*) \hspace{6.0truecm} \nonumber \\
=Q^{\rm tot} C \left\{\left[ \int_{z^*}^\infty (z_N-z^*)^2 e^{-\beta(z_N-z^*)^2} dz_N 
+  \int_{-\infty}^{-z^*} (z_N+z^*)^2 e^{-\beta(z_N-z^*)^2} dz_N \right] \right .\nonumber \\
 \left . + \left[ \int_{0}^{z^*} (z^*-z_0)^2 e^{-\beta(z^*-z_0)^2} dz_0
+\int_{-z^*}^0 (-z^*-z_0)^2 e^{-\beta(z^*-z_0)^2} dz_0 \right] \right\} \, .
\end{eqnarray}
The theoretical prediction of $R_{e,\perp}^2(z^*)$~\cite{Silberberg1982}:
\begin{eqnarray}
  R_{e,\perp}^2(z^*)
=\frac{1}{2\beta}\left\{1-4z^*\sqrt{\frac{\beta}{\pi}}+4{z^*}^2\beta \left[ 1-\textrm{erf}(z^*\sqrt{\beta})\right] \right\}
\end{eqnarray}




\subsubsection{Extended Silberberg's model in the thin film regime}
Total number of configurations of chains starting and terminating at $z^*$ is given by
\begin{eqnarray}
  Q^w(z_0=z^*,z_N) \hspace{11.0truecm} \nonumber \\
=Q^{\rm tot}C \left \{ \int_{z^*}^h e^{-\beta(z_N-z^*)^2} dz_N \qquad (k=0) \right .  \hspace{6.4truecm}  \nonumber \\
+\int_{-h}^{-z^*} e^{-\beta(z_N-z^*)^2} dz_N + \int_{h}^{2h-z^*} e^{-\beta(z_N-z^*)^2} dz_N \qquad (k=1) \hspace{2.85truecm} \nonumber \\
\left . +\int_{-2h+z^*}^{-h}e^{-\beta(z_N-z^*)^2} dz_N+\int_{2h+z^*}^{3h} e^{-\beta(z_N-z^*)^2} dz_N \qquad (k=2) \right . \hspace{2.5truecm}\nonumber \\
+ \qquad \cdots  \hspace{10.40truecm}  {\mathlarger{\mathlarger{\mathlarger {\mathlarger {\mathlarger {\mathlarger \}}}}}}}
\end{eqnarray}
and
\begin{eqnarray}
  Q^w(z_0,z_N=z^*) \hspace{11.0truecm} \nonumber \\
=Q^{\rm tot}C \left\{ \int_{0}^{z^*} e^{-\beta(z^*-z_0)^2} dz_0 \qquad (k=0) \right . \hspace{6.4truecm}  \nonumber \\
+\int_{-z^*}^{0} e^{-\beta(z^*-z_0)^2} dz_0 + \int_{2h-z^*}^{2h} e^{-\beta(z^*-z_0)^2} dz_0\qquad (k=1) \hspace{3.55truecm} \nonumber \\
\left . +\int_{2h}^{2h+z^*}e^{-\beta(z^*-z_0)^2} dz_N+\int_{-2h}^{-2h+z^*} e^{-\beta(z^*-z_0)^2} dz_0\qquad (k=2) \right . \hspace{2.5truecm}
\nonumber \\
+ \qquad \cdots \hspace{10.4truecm} {\mathlarger{\mathlarger{\mathlarger {\mathlarger {\mathlarger {\mathlarger \}}}}}}}
\end{eqnarray}
respectively, where $k$ denotes the number of swaps, $k=0,\,1,\,2,\,\ldots\,,n$.

\begin{description}
\item[For $k=0$]
\begin{eqnarray}
  Q^w(z_0=z^*,z_N)+Q^w(z_0,z_N=z^*) \Rightarrow C Q^{\rm tot}\int_0^{h}e^{-\beta(z-z^*)^2} dz \hspace{3.0truecm} \nonumber \\
=Q^{\rm tot} \left[\textrm{erf}((h-z^*)\sqrt{\beta})+\textrm{erf}(z^*\sqrt{\beta})\right] \hspace{1.0truecm}
\end{eqnarray}
\item[For $k=1$]
\begin{eqnarray}
  Q^w(z_0=z^*,z_N)+Q^w(z_0,z_N=z^*) 
\Rightarrow CQ^{\rm tot} \left\{\int_{-h}^{0} e^{-\beta(z-z^*)^2} dz + \int_h^{2h} e^{-\beta(z-z^*)^2} dz  \right\} 
\hspace{1.0truecm} \nonumber \\
=Q^{\rm tot} \left[-\textrm{erf}(z^*\sqrt{\beta})+\textrm{erf}((h+z^*)\sqrt{\beta})+\textrm{erf}((2h-z^*)\sqrt{\beta})
-\textrm{erf}((h-z^*)\sqrt{\beta})\right] \hspace{1.6truecm}
\end{eqnarray}
\item[For $k=2$]
\begin{eqnarray}
Q^w(z_0=z^*,z_N)+Q^w(z_0,z_N=z^*) \hspace{9.2truecm} \nonumber \\
\Rightarrow CQ^{\rm tot} \left\{\int_{2h}^{3h} e^{-\beta(z-z^*)^2}dz+\int_{-2h}^{-h} e^{-\beta(z-z^*)^2}dz \right \}  
\hspace{6.7truecm}\nonumber \\
=Q^{\rm tot}\left[\textrm{erf}((3h-z^*)\sqrt{\beta})-\textrm{erf}((2h-z^*)\sqrt{\beta})
-\textrm{erf}((h+z^*)\sqrt{\beta})+\textrm{erf}((2h+z^*))\sqrt{\beta})\right] \hspace{1.3truecm}
\end{eqnarray}
\end{description}
Considering the total number of chains up to $n$ swaps
\begin{eqnarray}
  Q^w(z_0=z^*,z_N)+Q^w(z_0,z_N=z^*) \hspace{5.0truecm} \nonumber \\
=Q^{\rm tot}\left[\textrm{erf}(((n+1)h-z^*)\sqrt{\beta})
+\textrm{erf}((nh+z^*)\sqrt{\beta}) \right] \xrightarrow{n \rightarrow \infty} 2 Q^{\rm tot}
\end{eqnarray}

The theoretical prediction of $R^2_{e,\perp}(z^*)$~\cite{Sussman2014}:
\begin{eqnarray}
  R_{e,\perp}^2(z^*)
	&=& \frac{Q^{\rm tot} C ( I_0 +I_1 +I_2 + \ldots +I_n}{Q^w(z_0=z^*,z_N)+Q^w(z_0,z_N=z^*)} \nonumber \\
&=& 2 \sqrt{\frac{\beta}{\pi}} \frac{I_0+I_1+I_2 + \ldots +I_n}
{\textrm{erf}(((n+1)h-z^*)\sqrt{\beta}+\textrm{erf}((nh+z^*)\sqrt{\beta})}
\end{eqnarray}

\begin{eqnarray}
I_0&=&\int_0^h (z-z^*)^2  e^{-\beta (z-z^*)^2}dz =\int_{-z^*}^{h-z^*} x^2 e^{-\beta x^2} dx \nonumber \\
&=&\frac{1}{4}\sqrt{\frac{\pi}{\beta^3}}\left[\textrm{erf}((h-z^*)\sqrt{\beta})+\textrm{erf}(z^*\sqrt{\beta})\right]
 -\frac{1}{2\beta}\left[(h-z^*)e^{-\beta (h-z^*)^2}+z^* e^{-\beta {z^*}^2} \right ]
\end{eqnarray}

\bigskip
\begin{eqnarray}
I_1&=&\int_{-h}^{-z^*}(-z^*-z)^2 e^{-\beta(z-z^*)^2}dz+\int_{h}^{2h-z^*}((2h-z^*)-z)^2 e^{-\beta(z-z^*)^2} dz \nonumber \\
&&+ \int_{-z^*}^0 (z-(-z^*))^2 e^{-\beta(z-z^*)^2}dz + \int_{2h-z^*}^{2h} (z-(2h-z^*))^2 e^{-\beta(z-z^*)^2} dz \nonumber \\
&=& \frac{1}{4}\sqrt{\frac{\pi}{\beta^3}}\left[\textrm{erf}((h+z^*)\sqrt{\beta})-\textrm{erf}(z^*\sqrt{\beta})
 +\textrm{erf}(2(h-z^*)\sqrt{\beta}) -\textrm{erf}((h-z^*)\sqrt{\beta})\right] \nonumber \\
&& -\frac{1}{2\beta}\left[
(h+z^*)e^{-\beta(h+z^*)^2}-z^* e^{-\beta {z^*}^2}+2(h-z^*)e^{-4\beta(h-z^*)^2} -(h-z^*)e^{-\beta(h-z^*)^2} \right] \nonumber \\
&&+\frac{2z^*}{\beta}\left[e^{-\beta(h+z^*)^2}-e^{-\beta {z^*}^2} \right]+\frac{2(h-z^*)}{\beta}\left[e^{-4\beta(h-z^*)^2}
-e^{-\beta(h-z^*)^2} \right] \nonumber \\
&& +2{z^*}^2\sqrt{\frac{\pi}{\beta}}\left[\textrm{erf}((h+z^*)\sqrt{\beta})-\textrm{erf}(z^*\sqrt{\beta}) \right] 
+ 2(h-z^*)^2\sqrt{\frac{\pi}{\beta}}\left[\textrm{erf}(2(h-z^*)\sqrt{\beta})-\textrm{erf}((h-z^*)\sqrt{\beta})\right]
\end{eqnarray}
\bigskip
\begin{eqnarray}
I_2&=& \int_{-2h+z^*}^{-h} (z-(-2h+z^*))^2 e^{-\beta(z-z^*)^2}dz +\int_{2h+z^*}^{3h} (z-(2h+z^*))^2 dz \nonumber \\
&&+\int_{2h}^{2h+z^*} (z-2h)^2 e^{-\beta(z-z^*)^2} dz+\int_{-2h}^{-2h+z^*} (z-(-2h+z^*))^2 e^{-\beta(z-z^*)^2} dz \nonumber \\
&=& \frac{1}{4}\sqrt{\frac{\pi}{\beta^3}}\left[\textrm{erf}((3h-z^*)\sqrt{\beta})-\textrm{erf}((2h-z^*)\sqrt{\beta})
-\textrm{erf}((h+z^*)\sqrt{\beta})+\textrm{erf}((2h+z^*)\sqrt{\beta}) \right] \nonumber \\
&&-\frac{1}{2\beta}\left[(3h-z^*)e^{-\beta(3h-z^*)^2}-(2h-z^*)e^{-\beta(2h-z^*)^2} 
-(h+z^*)e^{-\beta(h+z^*)^2}
   +(2h+z^*)e^{-\beta(2h+z^*)^2} \right] \nonumber \\
&&+ 2\frac{h}{\beta} \left[e^{-\beta(3h-z^*)^2}-e^{-\beta(2h-z^*)^2}
-e^{-\beta(h+z^*)^2}+e^{-\beta(2h+z^*)^2} \right] \nonumber \\
&& + 2h^2 \sqrt{\frac{\pi}{\beta}} \left[\textrm{erf}((3h-z^*)\sqrt{\beta})-\textrm{erf}((2h-z^*)\sqrt{\beta}) 
  -\textrm{erf}((h+z^*)\sqrt{\beta}) +\textrm{erf}((2h+z^*)\sqrt{\beta})\right]
\end{eqnarray}

\newpage
\section{Estimates of $T_g$ from the change of total potential energy $U_{\rm tot}(T)$}
\begin{figure*}[ht!]
\begin{center}
(a)\includegraphics[width=0.28\textwidth,angle=270]{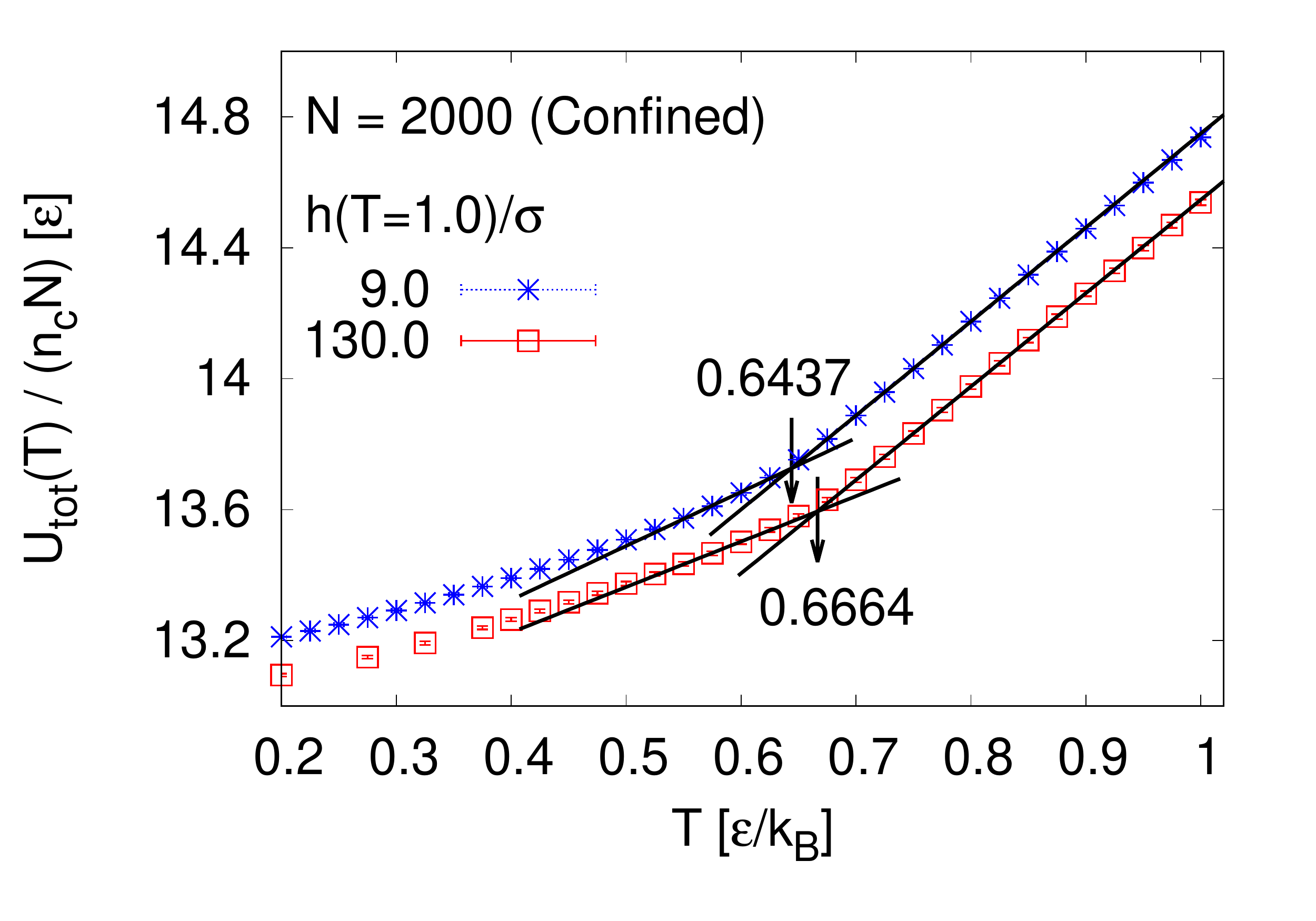}\hspace{0.5cm}
(b)\includegraphics[width=0.28\textwidth,angle=270]{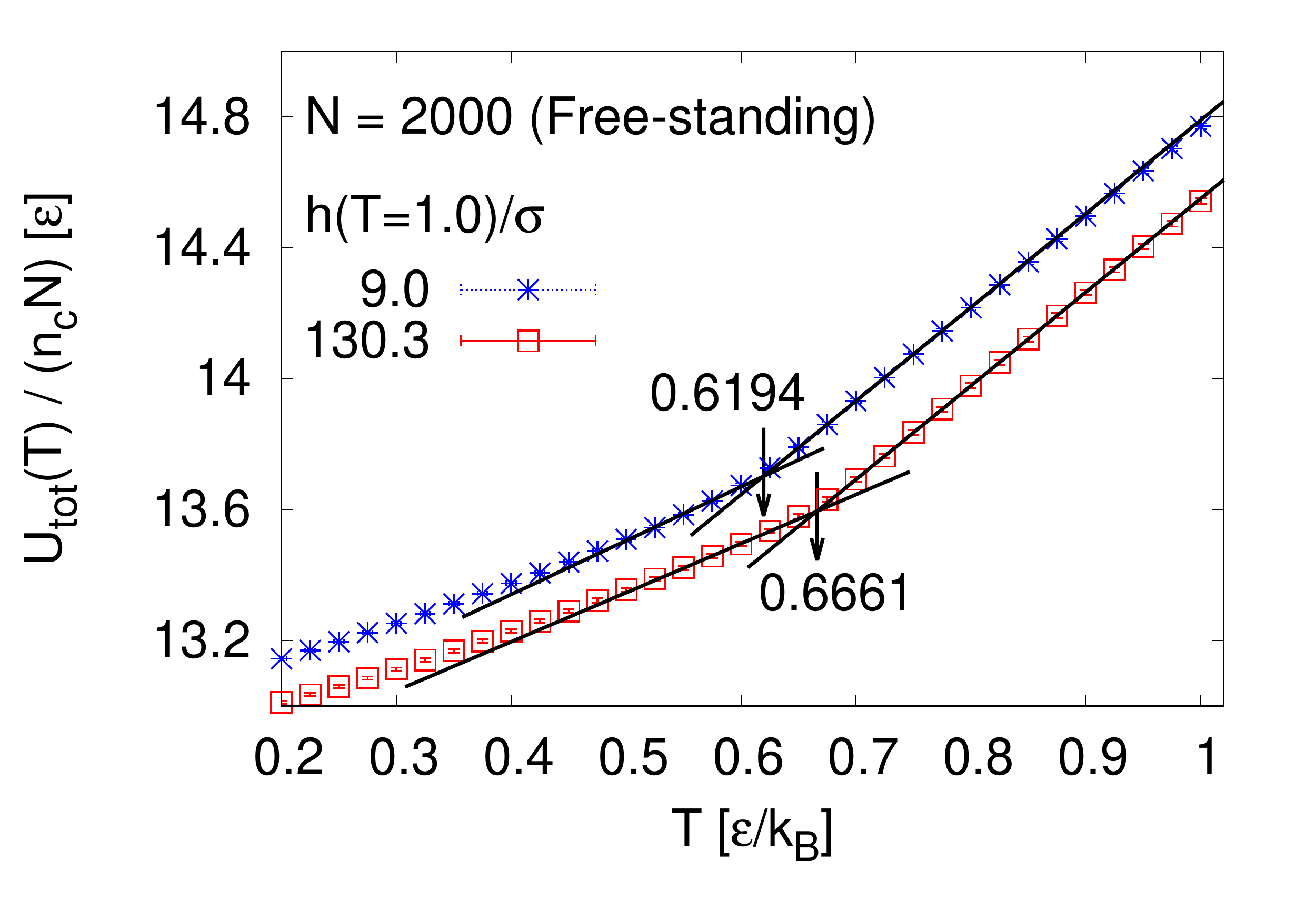}
\caption{Total potential energy $U_{\rm tot}(T)$ per monomer of confined (a) and free-standing polymer films (b) of two selected film thicknesses $h/R^{(0)}(N=2000)\approx 4.3$ and $0.3$, plotted as a function of $T$. For each set of data, $T_g$ indicated by an arrow is determined by the intersection between two linear fits, liquid branch ($U_{\rm tot}(T)=a_{\rm liquid}+\alpha_{\rm liquid}T$) and glass branch ($U_{\rm tot}(T)=a_{\rm glass}+\alpha_{\rm glass}T$). Each film contains $n_c=1000$ chains of $N=2000$ monomers. The same estimate of $T_g$ for bulk samples 
gives $T_g^{(0)}(N=2000)=0.6662(128)$ and is compatible with the results from the density $\rho(T)$ change.}

\label{fig-Tg-Utot-film}
\end{center}
\end{figure*}


%
\end{document}